%% file: paper_sharc_2023.tex
\documentclass[conference, transmag]{IEEEtran}
\IEEEoverridecommandlockouts
\usepackage{cite}
\usepackage{amsmath,amssymb,amsfonts}
\usepackage{algorithmic}
\usepackage{graphicx}
\usepackage{textcomp}
\usepackage{xcolor}
\usepackage{glossaries}
\usepackage{tikz}
\usepackage{pgfplots}
\usepackage{siunitx}

\pgfplotsset{compat=1.8}

\graphicspath{{fig/}}

\newacronym{dss}{DSS}{distributed satellite system}
\newacronym{sdr}{SDR}{software-defined radio}
\newacronym{cots}{COTS}{commercial off-the-shelf}
\newacronym{rx}{RX}{receive}
\newacronym{tx}{TX}{transmit}
\newacronym{trx}{TRX}{transceiver}
\newacronym{soc}{SoC}{System-on-a-Chip}
\newacronym{fpga}{FPGA}{field-programmable gate array}
\newacronym{pl}{PL}{programmable logic}
\newacronym{ps}{PS}{processing system}
\newacronym{rf}{RF}{radio-frequency}
\newacronym{twtt}{TWTT}{two-way time transfer}
\newacronym{tof}{ToF}{time of flight}
\newacronym{toa}{ToA}{time of arrival}
\newacronym{hdl}{HDL}{hardware description language}
\newacronym{gnss}{GNSS}{global navigation satellite system}
\newacronym{ecb}{ECB}{equivalent complex baseband}
\newacronym{crlb}{CRLB}{Cramer-Rao Lower Bound}
\newacronym{snr}{SNR}{Signal-to-Noise Ratio}
\newacronym{dqpsk}{DQPSK}{differential quadrature phase-shift keying}
\newacronym{pa}{PA}{power amplifier}
\newacronym{lna}{LNA}{low-noise amplifier}
\newacronym{rssi}{RSSI}{received signal strength indicator}
\newacronym{sff}{SFF}{satellite formation flying}

\def\BibTeX{{\rm B\kern-.05em{\sc i\kern-.025em b}\kern-.08em
    T\kern-.1667em\lower.7ex\hbox{E}\kern-.125emX}}

\newcommand{\figref}[1]{Fig. \ref{#1}}

\begin{document}

\title{Towards Wireless Ranging and Synchronization using CubeSat Software-Defined Radio Subsystems\\
}

\author{\IEEEauthorblockN{Markus Gardill}
\IEEEauthorblockA{\textit{Chair for Electronic Sensors and Systems} \\
\textit{Brandenburg University of Technology}\\
Cottbus, Germany \\
markus.gardill.de@ieee.org}
\and
\IEEEauthorblockN{Dominik Pearson}
\IEEEauthorblockA{\textit{Center for Telematics}\\
Würzburg, Germany \\
dominik.pearson@telematik-zentrum.de}
\and
\IEEEauthorblockN{Julian Scharnagl}
\IEEEauthorblockA{\textit{Center for Telematics}\\
Würzburg, Germany \\
julian.scharnagl@telematik-zentrum.de}
\and
\IEEEauthorblockN{ }
\IEEEauthorblockA{ }
\\
\and
\IEEEauthorblockN{Klaus Schilling}
\IEEEauthorblockA{\textit{Center for Telematics}\\
Würzburg, Germany \\
klaus.schilling@telematik-zentrum.de}
}

\maketitle

\begin{abstract}
An approach towards wireless ranging and synchronization using commercial of-the-shelf software-defined radio payloads for small satellites, esp. CubeSats is studied.
The approach only relies on the programmable logic configuration and processing system software.
No hardware modifications or additions to the payloads are necessary.
Experimental evaluation of the initial implementation shows a standard deviation of time-of-flight-based ranging measurements in the order of $\SI{1}{\centi\meter}$, which renders the concept of interest for distributed satellite system missions.
\end{abstract}


\section{Introduction}
Cooperating satellites allow for new application areas, improved mission performance and cost savings and thus receive increasing interest in recent years.

Several manifestations of this concept have been reported, ranging from \glspl{dss}, e.g., for improved communications \cite{saeed2020} or \gls{sff} for Earth observation \cite{bachmann2021, schilling2015, Draschka2021}, up to the concept of Fractionated Spacecraft \cite{sweeting2018}.
Depending on the mission requirements, precise synchronization and ranging are necessary between individual satellites, especially for missions with very high relative positioning or synchronization requirements that cannot be met by typical \gls{gnss} receivers for small satellites.
A recent overview on synchronization for \gls{dss} is presented in \cite{marrero2022}.

While the \gls{dss} concept has been studied for several classes of satellites, it is of particular interest for small satellites, often implemented according to the CubeSat standard \cite{nasa2020}.
References \cite{SCHARNAGL2022580, VonArnim2022} indeed describe CubeSat-based \gls{dss} missions, but high-precision wireless synchronization as, e.g., used in large-scale missions such as \cite{bachmann2021}, is typically not applied.


A method for wireless sub-nanosecond synchronization for terrestrial \gls{sdr} networks has, e.g., recently been proposed in \cite{prager2020}, achieving sub nanosecond synchronization for phase-coherent distributed systems.
The authors achieve excellent results for multiple nodes, but rely on a pre-synchronization, e.g., using \glspl{gnss}.
Based on the concept from \cite{prager2020}, in this work we want to study how the method can be applied to \gls{cots} CubeSat \gls{sdr} payloads, without requiring any additional pre-synchronization.

\section{CubeSat Software-Defined Radio Architecture}
\begin{figure}[t]
    \scriptsize
    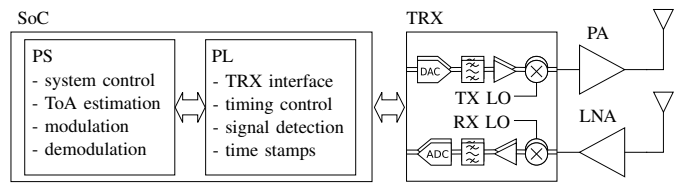
    \caption{
        Typical architecture of \gls{sdr} for CubeSat applications.
        A fully integrated front-end is interfaced to a \gls{soc} including \gls{pl} and \gls{ps} resources.
        The mentioned bullet points indicate how wireless ranging and synchronization proposed in this work is realized by functions implemented in the \gls{pl} and \gls{ps}.
        }
    \label{fig:cubesat_sdr_architecture}
\end{figure}

\glspl{sdr} are increasingly being used in CubeSats \cite{nasa2020}.
Their architecture often consists of an integrated \gls{trx} together with a \gls{soc}, i.e., a combination of \gls{fpga} \gls{pl} and a general-purpose processor-based \gls{ps}, as illustrated in \figref{fig:cubesat_sdr_architecture}.

Specific realizations are, e.g., based on Analog Devices AD936x \glspl{trx} together with Xilinx Zynq-7000 family \glspl{soc}.
Several \gls{cots} CubeSat \glspl{sdr} are available implementing this architecture, e.g. refer to the products from \cite{sdr_alenspace_2022, sdr_gomspace_2022}.

In this work, we consequently consider how precise ranging and synchronization can be implemented using this architecture, i.e., only as a software configuration option to be easily integrated in an \gls{sdr}-equipped CubeSat mission.

\section{Two-Way Time Transfer}
A general idea for synchronization is \gls{twtt}.
It has a long heritage and has been reported in several works for satellite applications before \cite{kirchner1991}.
We hence only present the main ideas and our models in this work, for a more elaborate review the reader is e.g., referred to \cite{prager2020}.
We consider a linear clock model according to
\begin{equation}
    \tau_i = \alpha_i t + \phi_i,
    \label{eq:clock_model}
\end{equation}
where $\tau_i$ is the local clock of satellite $i \in \{\rm{A}, \rm{B}\}$, $t$ is the global clock, $\alpha_i$ is the clock skew (frequency difference) and $\phi_i$ is the clock offset.
It is assumed to be adequate for the relatively short synchronization intervals considered in this case study, although higher-order clock-models have been reported for improved accuracy\cite{xie2016}.

\begin{figure}
    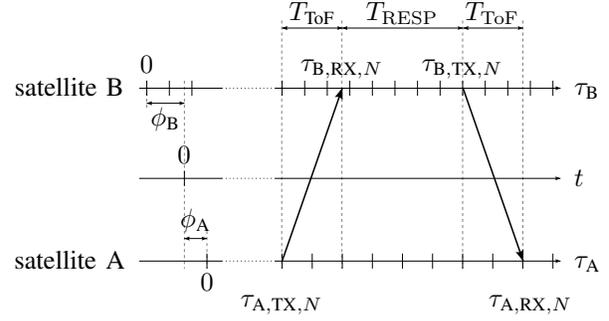
    \caption{Illustration of clock model and single \gls{twtt} measurement $N$.}
    \label{fig:twtt_principle}
\end{figure}

Three parameters are of interest to solve the synchronization and ranging challenge: the \gls{tof} $T_{\text{ToF}}$, which is related to the distance $d_{\rm{A}, \rm{B}}$ between both satellites and the velocity of propagation $c_0$ by $T_{\text{ToF}} = d_{\rm{A}, \rm{B}}/c_0 $, the time offset $\Delta T_{\rm{A},\rm{B}} = \phi_{\rm{B}} - \phi_{\rm{A}}$ between both satellites, and the relative clock skew $\alpha_{\text{A}}/\alpha_{\text{B}}$.
As illustrated in \figref{fig:twtt_principle}, the principle of \gls{twtt} is to estimate all those parameters by time-stamping the time instances of \gls{rx} and \gls{tx} of synchronization waveforms exchanged between the satellites, collecting this timing information at a single node (e.g., transmitting $\tau_{\rm{B},\rm{RX}}$ and $\tau_{\rm{B},\rm{TX}}$ to satellite A), and solving a set of equations.
If multiple \gls{twtt} measurements are used, we denote this by adding the index $N$ to all quantities.
If the focus is only on a single \gls{twtt} measurement, this index is dropped for convenience.

An initial estimate of the clock offset could be obtained by
\begin{equation}
    \Delta \hat{T'}_{\text{A},\text{B}} = \frac{(\tau_{\text{B}, \text{RX}} + \tau_{\text{B}, \text{TX}}) - (\tau_{\text{A}, \text{RX}} + \tau_{\text{A}, \text{TX}})}{2}.
    \label{eq:clock_offset_estimate_initial}
\end{equation}
However, by plugging the clock model from \eqref{eq:clock_model} in \eqref{eq:clock_offset_estimate_initial} it can be seen that $\Delta \hat{T'}_{\text{A},\text{B},N} = \Delta T_{\text{A},\text{B},N}$ only holds for identical clock-skews $\alpha_{\text{A}} = \alpha_{\text{B}}$.
Else an error term prevails in \eqref{eq:clock_offset_estimate_initial}.
One solution if a non-zero clock-skew exists is to use two successive \gls{twtt} measurements, because from $\Delta \hat{T'}_{\text{A},\text{B},N+1} - \Delta \hat{T'}_{\text{A},\text{B},N}$ it can be shown that
\begin{equation}
    \frac{\alpha_{\text{B}}}{\alpha_{\text{A}}} =
    \frac{2 (\Delta \hat{T'}_{\text{A},\text{B},N+1} - \Delta \hat{T'}_{\text{A},\text{B},N})}%
    {\tau_{\text{A},\text{TX},N+1} - \tau_{\text{A},\text{TX},N} + \tau_{\text{A},\text{RX},N+1} - \tau_{\text{A},\text{RX},N}}  + 1.
    \label{eq:relative_clock_skew_estimate}
\end{equation}
Once \eqref{eq:relative_clock_skew_estimate} has been determined the \gls{tof} can be estimated by a single \gls{twtt} measurement using
\begin{equation}
    \hat{T}_{\text{ToF}} = \frac{\tau_{\text{A},\text{RX}} - \tau_{\text{A},\text{TX}}}{2} - \frac{\alpha_{\text{B}}}{\alpha_{\text{A}}} \frac{\tau_{\text{B},\text{TX}} - \tau_{\text{B},\text{RX}}}{2}.
    \label{eq:tof_estimate}
\end{equation}
A corrected clock-offset estimate is then given by
\begin{multline}
    \Delta \hat{T}_{\rm{A},\rm{B}} = \frac{(\tau_{\text{B},\text{RX}} + \tau_{\text{B},\text{TX}})}{2}\\
    - \left( \tau_{\text{A},\text{TX}} + \frac{\alpha_{\text{B}}}{\alpha_{\text{A}}} \frac{\tau_{\text{A},\text{RX}} - \tau_{\text{A},\text{TX}}}{2} \right).
    \label{eq:clock_offset_estimate}
\end{multline}
Note that \eqref{eq:clock_offset_estimate} is based on a transformation which sets $t_{\text{A},\text{RX},N} = 0$, i.e., the current \gls{twtt} measurement starts at zero global time.
This is equivalent to $\tau_{\text{A},\text{RX},N} = \alpha_{\text{A}} t_{\text{A},\text{RX},N} + \phi_{\text{A}} = \phi_{\text{A}}$.
Clearly, the framework of equations \eqref{eq:relative_clock_skew_estimate}, \eqref{eq:tof_estimate}, and \eqref{eq:clock_offset_estimate} allows to solve the synchronization challenge by determining $\tau_{\text{A},\text{TX},n}$, $\tau_{\text{B},\text{RX},n}$, $\tau_{\text{B},\text{TX},n}$, $\tau_{\text{A},\text{RX},n}$ from two successive \gls{twtt} measurements $n \in {N, N+1}$.

\section{Logic Implementation}
\begin{figure}
    \scriptsize
    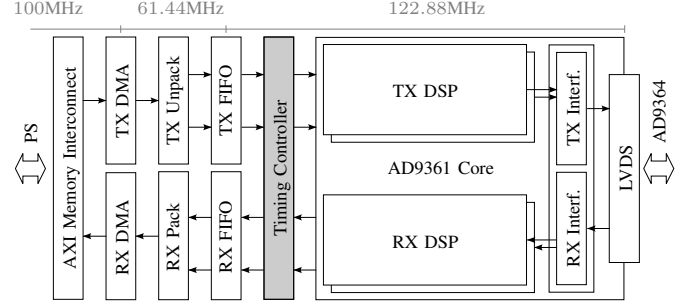
    \caption{
        \Gls{hdl} design of \gls{pl} used in this work including clock domains.
        The Timing Controller implements the synchronization logic developed in this work, the remaining blocks are based on a publicly available reference design.}
    \label{fig:extended_hdl_reference_design}
\end{figure}

For the basic implementation of an AD936x-based \gls{sdr} together with a Zynq-based \gls{soc}, \gls{hdl} reference designs are available from Analog Devices \cite{adi_hdl_reference_designs}.
However, the reference design basically performs \gls{tx} and \gls{rx} via buffers, which are filled and read by the \gls{ps}.
To obtain the required time-stamped measurements for \gls{twtt}, two additional mechanisms need to be implemented in the \gls{pl}: the \gls{tx} timing must be precisely controlled, such that $\tau_{i,\text{TX},n}$, $i \in \{\text{A},\text{B}\}$ are precisely known in the local time base and the \gls{rx} \gls{toa} must be estimated such that $\tau_{i,\text{RX},n}$, $i \in \{\text{A},\text{B}\}$ are determined.
Therefore, the \gls{hdl} reference design has been extended as illustrated in \figref{fig:extended_hdl_reference_design}.

To achieve precisely controlled \gls{tx} timing, the basic idea of our implementation is to fill the \gls{tx} buffer as usual, but block the data path at some point until the desired \gls{tx} time stamp $\tau_{i,\text{TX},n}$, $i \in \{\text{A},\text{B}\}$ is reached by the system's clock.
To obtain the time stamps $\tau_{i,\text{RX},n}$, $i \in \{\text{A},\text{B}\}$ in the \gls{rx} direction, the \gls{rx} data stream is continuously monitored by the timing controller.
Only if a signal is detected by the \gls{pl}, the \gls{rx} buffer is filled, a time stamp is added to the first sample of the \gls{rx} buffer and the buffer is processed by the \gls{ps}.
Signal detection is currently based on a \gls{rssi} threshold, but more robust methods such as the use of correlation-based preamble detection could be easily implemented.
Via this single timing reference in the \gls{rx} buffer, the precise $\tau_{i,\text{RX},n}$, $i \in \{\text{A},\text{B}\}$ can be determined by a correlation-based approach in the \gls{ps}, exploiting the characteristics of the synchronization waveforms.

\begin{figure}
    \scriptsize
    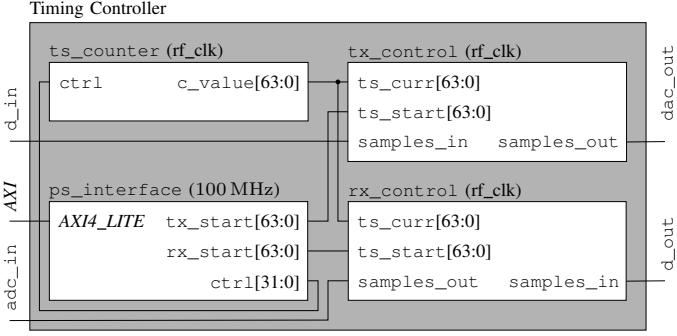
    \caption{
        Simplified structure of timing controller IP-core.
        While several additional details are hidden, note in particular that most blocks are clocked at the RF clock $f_{\mathrm{rf\_clk}} = 2 \cdot f_{\text{s}}$.
        The \gls{ps} interface operates at \SI{100}{\mega\hertz} and contains additional logic to ensure correct clock-domain crossing.
        }
    \label{fig:timing_controller_hdl_reference_design}
\end{figure}

The simplified \gls{hdl} structure of the specific Timing Controller implementation is illustrated in \figref{fig:timing_controller_hdl_reference_design}.
The \texttt{ps\_interface} module implements the AXI4-LITE interface and is used to control the entire Timing Controller IP-Core.
All parameters required to control the IP-Core can be set through register writes from the PS. Status bits from all the modules are collected in one register and can be read by the PS.

The \texttt{ts\_counter} module provides the timing reference both for the \texttt{tx\_control} and \texttt{rx\_control} modules.
It does so by counting the rising clock edges of the \texttt{rf\_clk} clock and outputs this value in the $64 \; \mathrm{Bit}$ signal \texttt{c\_value}.
The timing control of the transmit path is implemented by the \texttt{tx\_control} module.
It is achieved by interfering with the data interface between the FIFO buffers (\texttt{d\_in}) and AXI9361 (\texttt{dac\_out}).
The current timer value provided by the \texttt{ts\_counter} module is compared against the timestamp of the desired start of transmission.
The signal of the comparator output feeds a multiplexer that either ties down the data path signals or connects the signals between the FIFO buffer directly to the signals of the AXI9361 core, without modifying them.

The \texttt{rx\_control} module implements all features related to the timing control of the \gls{rx} path.
It supports two modes of operation.
As illustrated in \figref{fig:timing_controller_hdl_reference_design}, it can either start a reception triggered by reaching a desired timestamp \texttt{tx\_start}, just as the \texttt{tx\_control} module does for transmissions.
However, it also provides a mode of operation where receptions are triggered by exceeding a conﬁgurable signal power value. Therefore, the \texttt{rx\_control} module includes logic to calculate the \emph{RSSI} of the received signal in real time.
In both modes, the timestamp \texttt{ts\_start} of the first accepted sample is written to a register, which can be read by the \gls{ps}.

\section{Signal Processing}
Like the method proposed in \cite{prager2020}, chirp signals are used as synchronization waveforms.
The \gls{ecb} representation of the chirp signal $s(t)$ is given by
\begin{equation}
    s(t) =
    \begin{cases}
        \exp \left(\jmath 2 \pi \left( \frac{B_{\text{c}} f_{\text{s}}}{2 l_{\text{c}}} t - \frac{B_{\text{c}}}{2}\right) t \right) & \text{for } 0 \le t \le T_{\text{c}}\\
        0
    \end{cases},
    \label{eq:chirp_waveform}
\end{equation}
where $B_{\text{c}}$ is the bandwidth of the Chirp, $f_{\text{s}}$ the sampling rate, $l_{\text{c}}$ the length of the chirp in samples, and $T_{\text{c}} = l_{\text{c}}/f_{\text{s}}$ the duration of the chirp.
Since the chirp is generated in the time domain $\tau_{\text{A}}$ of satellite $A$, the \gls{ecb} signal $r_\mathrm{B}(\tau_\mathrm{B})$ received by satellite $\text{B}$ is subject to the time transformation $\tau_{\text{A}} \rightarrow \tau_{\text{B}}$.
It then appears as
\begin{multline}
	\label{eq:rxsignaldB2}
	r_\mathrm{B}(\tau_\mathrm{B}) = s\left(\frac{\alpha_\mathrm{A}}{\alpha_\mathrm{B}}(\tau_\mathrm{B}-\tau_{\text{B},\text{RX}})\right)
	e^{\mathrm{j}2\pi  f_\mathrm{err} \tau_\mathrm{B}}\\
	e^{-\mathrm{j}2\pi f_\mathrm{c}\left(\alpha_\mathrm{B}^{-1}\alpha_\mathrm{A} \left(	\phi_\mathrm{B}+\tau_{B_\mathrm{ToF}}\right)-\phi_\mathrm{A}\right)}
	e^{\gamma_\mathrm{A,B}^\mathrm{err}},
\end{multline}
where $f_\mathrm{err}$ is the frequency and $\gamma_\mathrm{A,B}^\mathrm{err}$ the phase difference between both satellites, and $\tau_{B_\mathrm{ToF}}$ is the \gls{tof} expressed in the clock domain of satellite $\text{B}$.
To estimate the \gls{toa}, satellite $\text{B}$ generates the chirp \eqref{eq:chirp_waveform} in its own time base and performs the discrete-time equivalent of the correlation $d_{\text{A}, \text{A}}(\tau'_B) = (s\star r_\mathrm{B} )(\tau'_{\text{B}})$ with the received signal ($\tau'_{\text{B}}$ is used to denote the correlation lags).
A peak detection followed by peak interpolation based on a sinc nonlinear least-squares is then used to estimate $\hat{\tau}_{\text{B},\text{RX}} = \mathrm{argmax}_{\tau'_{\text{B}}} |d_{\text{A}, \text{B}}(\tau'_B)|$ with sub-sample accuracy from the discrete-time correlation result.

Exchange of the timestamps between the stations is realized by a simple \gls{dqpsk} scheme implemented in the \gls{ps}.

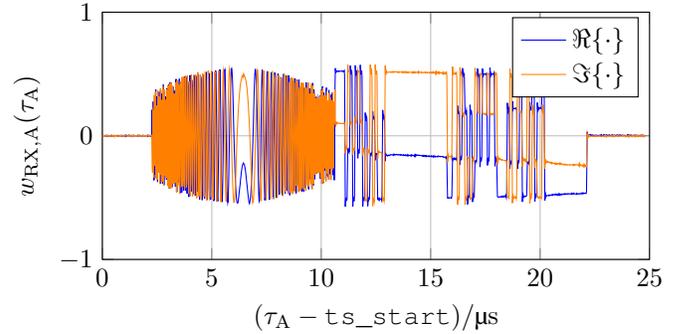
\begin{figure}[ht]
    \begin{tikzpicture}
        \begin{axis}[%
        x filter/.code={\pgfmathdivide{#1}{1e-6}},
        xmin = 0, xmax = 25,
        ymin = -1, ymax = 1,
        ytick distance = 1,
        grid = both,
        major grid style = {lightgray},
        minor grid style = {lightgray!40},
        width = \columnwidth,
        height = 0.55\columnwidth,
        xlabel = {$(\tau_\mathrm{A}-\mathrm{\texttt{ts\_start}})/\si{\micro\second}$},
        ylabel = {$w_\mathrm{RX,A}(\tau_\mathrm{A})$},
        legend cell align=left,
        yticklabel style={
            /pgf/number format/fixed,
            /pgf/number format/precision=2},
        xticklabel style={
            /pgf/number format/fixed,
            /pgf/number format/precision=2}
            ]
         
   \addplot[mark=none, color = blue] table[x=t, y=real] {data/signal.dat};
   \addplot[mark=none, color = orange] table[x=t, y=imag] {data/signal.dat};

    \legend{$\Re\{\cdot\}$, $\Im\{\cdot\}$}
   
    \end{axis}
        
    \end{tikzpicture}
    \caption{
        Complete \gls{twtt} baseband waveform recorded at node A consisting of received chirp signal $r_\mathrm{A}(\tau_\mathrm{A})$ and \gls{dqpsk}-modulated timestamps.
        $\Re\{\cdot\}$ refers to the real part, $\Im\{\cdot\}$ to the imaginary part of the \gls{ecb} signal.}
    \label{fig:twtt_waveform}
\end{figure}

A complete baseband waveform $w(t)$ as used for the \gls{twtt} is shown in \figref{fig:twtt_waveform}.
The chirp signal has a bandwidth of $B_{\textrm{c}} = \SI{38}{\mega\hertz}$, is sampled at $f_\mathrm{s}=\SI{61.44}{\mega\hertz}$ and has a length of $l_\mathrm{c}=512$. The duration therefore is $T_{\textrm{c}} = l_{\text{c}}/f_{\text{s}} = \SI{8.3}{\micro\s}$, resulting in a sweeprate of $B_{\textrm{c}}/T_{\textrm{c}} = \SI{4.56e12}{\hertz\per\s}$.
The \gls{dqpsk} part contains eight status bits, the \gls{tx} timestamp ($\tau_\mathrm{B,TX}$) encoded with 64 bits as well as the \gls{rx} timestamp ($\tau_\mathrm{B,RX}$) encoded with 128 bits.
The waveform was recorded at node A during a \gls{twtt} measurement that was performed using antennas as described in section \ref{eval}.

\section{Experimental Evaluation}
\label{eval}
The experimental evaluation was done using development kits based on an Avnet Zedboard with AD-FMCOMMS4-EBZ module, according to recommendations for the CubeSat \gls{sdr} platform from \cite{sdr_alenspace_2022}.
Two of those evaluation kits were used in different configurations during the evaluation:
a) connected by cables and b) by using $\SI{7}{dBi}$ antennas connected to the \gls{rf} ports of the FMCOMMS4 modules at a device distance of about $\SI{1.8}{\meter}$.

\begin{figure}[ht]
    \begin{tikzpicture}
        \begin{axis}[%
        x filter/.code={\pgfmathdivide{#1}{1e6}},
        y filter/.code={\pgfmathdivide{#1}{1e-2}},
        ymin = 0, ymax = 4,
        ytick distance = 1,
        grid = both,
        major grid style = {lightgray},
        minor grid style = {lightgray!40},
        width = \columnwidth,
        height = 0.55\columnwidth,
        xlabel = {$B_\mathrm{c}/\si{\mega\hertz}$},
        ylabel = {$\sigma_\mathrm{ToF} \cdot c_0/\si{\centi\meter}$},
        legend cell align=left,
        yticklabel style={
            /pgf/number format/fixed,
            /pgf/number format/precision=2},
        xticklabel style={
            /pgf/number format/fixed,
            /pgf/number format/precision=2}
            ]
         
   \addplot[mark=none, color = blue] table[x=value,y index = 6] {data/dev2_antenna/3ghz2/ch_ll_256_bw.dat};
   \addplot[mark=none, color = green] table[x=value,y index = 6] {data/dev2_antenna/3ghz2/ch_ll_512_bw.dat};
   \addplot[mark=none, color = orange] table[x=value,y index = 6] {data/dev2_antenna/3ghz2/ch_ll_768_bw.dat};
   \addplot[mark=none, color = red] table[x=value,y index = 6] {data/dev2_antenna/3ghz2/ch_ll_1024_bw.dat};
   \addplot[mark=none, color = purple] table[x=value,y index = 6] {data/dev2_antenna/3ghz2/ch_ll_1280_bw.dat};

   \addplot[dashed, mark=none, color = blue] table[x=value,y index = 6] {data/dev2_cable/3ghz2/ch_ll_256_bw.dat};
   \addplot[dashed, mark=none, color = green] table[x=value,y index = 6] {data/dev2_cable/3ghz2/ch_ll_512_bw.dat};
   \addplot[dashed, mark=none, color = orange] table[x=value,y index = 6] {data/dev2_cable/3ghz2/ch_ll_768_bw.dat};
   \addplot[dashed, mark=none, color = red] table[x=value,y index = 6] {data/dev2_cable/3ghz2/ch_ll_1024_bw.dat};
   \addplot[dashed, mark=none, color = purple] table[x=value,y index = 6] {data/dev2_cable/3ghz2/ch_ll_1280_bw.dat};

    \addplot[color = black] table[x=bw,y = 256] {data/crlb.dat};
    \addplot[color = black] table[x=bw,y = 1280] {data/crlb.dat};

    \legend{ 256, 512, 768, 1024, 1280}
   
    \end{axis}
        
    \end{tikzpicture}
    \caption{
        Measured standard deviation for estimation of $T_{\text{ToF}}$ using two experimental platforms vs. chirp bandwidth, evaluated for different chirp lengths.
        Dashed lines show cable-connected configuration, solid lines show wireless configuration.
        Solid black lines are \gls{crlb} for chirp lengths of $256$ (upper line) and $1280$ (lower line) evaluated at an \gls{snr} of \SI{30}{dB}.}
    \label{fig:variance_tof_measurement}
\end{figure}
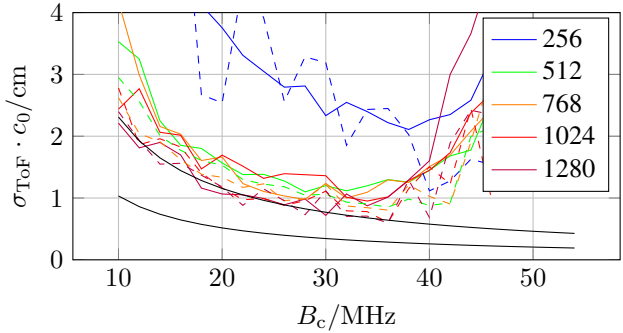

To study how the waveform parameters influence the precision of the \gls{tof} measurement, the variance $\sigma_\mathrm{ToF}$ of estimating $T_{\text{ToF}}$ in the clock domain of system A was estimated using $n_{\text{trial}} = 1000$ \gls{twtt} measurements for different $B_{\text{c}}$ and $l_{\text{c}}$.
The results are presented in \figref{fig:variance_tof_measurement}.

It can be seen that the measured results follow the trend modeled by the \gls{crlb} up to a bandwidth of about $B_\mathrm{c} \approx \SI{40}{\mega\hertz}$.
The reason for the starting deviation at this frequency has not entirely been determined, but is attributed to a general degradation of the sinc nonlinear least-squares peak interpolation when $B_\mathrm{c}$ approaches the system sampling rate of $f_{\text{s}} = \SI{61.44}{\mega\hertz}$, which was also observed in simulations.
A precision (in terms of standard deviation) of around $\sigma_\mathrm{ToF} \cdot c_0 \approx \SI{1}{\centi\meter}$ could be realized at a bandwidth of $B_{\text{c}}=\SI{36}{\mega\hertz}$.

\section{Conclusion}
We have shown in this work, that wireless synchronization and ranging can be achieved with state-of-the-art \gls{sdr} payload architectures for small satellites and even CubeSats.
The entire mechanism is implemented in the \gls{pl} and \gls{ps} of the system and does not require any external hardware or pre-sychronization.
Since the current approach was evaluated in a lab setup, future work is focused on the integration of the experimental evaluation platform with external devices such as \gls{pa} and \gls{lna} for a realistic evaluation of the approach considering satellite distances of up to 100's of \si{\kilo\meter}.

\bibliographystyle{IEEEtran}
\bibliography{references}

\end{document}

%% file: fig/sdr_architecture.pdf_tex
\begingroup%
  \makeatletter%
  \providecommand\color[2][]{%
    \errmessage{(Inkscape) Color is used for the text in Inkscape, but the package 'color.sty' is not loaded}%
    \renewcommand\color[2][]{}%
  }%
  \providecommand\transparent[1]{%
    \errmessage{(Inkscape) Transparency is used (non-zero) for the text in Inkscape, but the package 'transparent.sty' is not loaded}%
    \renewcommand\transparent[1]{}%
  }%
  \providecommand\rotatebox[2]{#2}%
  \newcommand*\fsize{\dimexpr\f@size pt\relax}%
  \newcommand*\lineheight[1]{\fontsize{\fsize}{#1\fsize}\selectfont}%
  \ifx\svgwidth\undefined%
    \setlength{\unitlength}{252.28346457bp}%
    \ifx\svgscale\undefined%
      \relax%
    \else%
      \setlength{\unitlength}{\unitlength * \real{\svgscale}}%
    \fi%
  \else%
    \setlength{\unitlength}{\svgwidth}%
  \fi%
  \global\let\svgwidth\undefined%
  \global\let\svgscale\undefined%
  \makeatother%
  \begin{picture}(1,0.29213483)%
    \lineheight{1}%
    \setlength\tabcolsep{0pt}%
    \put(0,0){\includegraphics[width=\unitlength,page=1]{sdr_architecture.pdf}}%
    \put(0.88176333,0.23595506){\color[rgb]{0,0,0}\makebox(0,0)[t]{\lineheight{1.25}\smash{\begin{tabular}[t]{c}PA\end{tabular}}}}%
    \put(0.88176333,0.11235955){\color[rgb]{0,0,0}\makebox(0,0)[t]{\lineheight{1.25}\smash{\begin{tabular}[t]{c}LNA\end{tabular}}}}%
    \put(0.59552126,0.25842697){\color[rgb]{0,0,0}\makebox(0,0)[lt]{\lineheight{1.25}\smash{\begin{tabular}[t]{l}TRX\end{tabular}}}}%
    \put(0,0){\includegraphics[width=\unitlength,page=2]{sdr_architecture.pdf}}%
    \put(0.01425356,0.25842697){\color[rgb]{0,0,0}\makebox(0,0)[lt]{\lineheight{1.25}\smash{\begin{tabular}[t]{l}SoC\end{tabular}}}}%
    \put(0,0){\includegraphics[width=\unitlength,page=3]{sdr_architecture.pdf}}%
    \put(0.03694723,0.20176082){\color[rgb]{0,0,0}\makebox(0,0)[lt]{\lineheight{1.25}\smash{\begin{tabular}[t]{l}PS\end{tabular}}}}%
    \put(0.30468651,0.20176082){\color[rgb]{0,0,0}\makebox(0,0)[lt]{\lineheight{1.25}\smash{\begin{tabular}[t]{l}PL\end{tabular}}}}%
    \put(0.30608199,0.1652093){\color[rgb]{0,0,0}\makebox(0,0)[lt]{\lineheight{1.25}\smash{\begin{tabular}[t]{l}- TRX interface\\- timing control\\- signal detection\\- time stamps\\\\\end{tabular}}}}%
    \put(0.03641908,0.16610791){\color[rgb]{0,0,0}\makebox(0,0)[lt]{\lineheight{1.25}\smash{\begin{tabular}[t]{l}- system control\\- ToA estimation\\- modulation\\- demodulation\\\end{tabular}}}}%
    \put(0,0){\includegraphics[width=\unitlength,page=4]{sdr_architecture.pdf}}%
    \put(0.66836222,0.1389703){\color[rgb]{0,0,0}\makebox(0,0)[lt]{\lineheight{1.25}\smash{\begin{tabular}[t]{l}TX LO\end{tabular}}}}%
    \put(0.66560342,0.10357953){\color[rgb]{0,0,0}\makebox(0,0)[lt]{\lineheight{1.25}\smash{\begin{tabular}[t]{l}RX LO\end{tabular}}}}%
    \put(0,0){\includegraphics[width=\unitlength,page=5]{sdr_architecture.pdf}}%
  \end{picture}%
\endgroup%

%% file: fig/twtt_principle.pdf_tex
\begingroup%
  \makeatletter%
  \providecommand\color[2][]{%
    \errmessage{(Inkscape) Color is used for the text in Inkscape, but the package 'color.sty' is not loaded}%
    \renewcommand\color[2][]{}%
  }%
  \providecommand\transparent[1]{%
    \errmessage{(Inkscape) Transparency is used (non-zero) for the text in Inkscape, but the package 'transparent.sty' is not loaded}%
    \renewcommand\transparent[1]{}%
  }%
  \providecommand\rotatebox[2]{#2}%
  \newcommand*\fsize{\dimexpr\f@size pt\relax}%
  \newcommand*\lineheight[1]{\fontsize{\fsize}{#1\fsize}\selectfont}%
  \ifx\svgwidth\undefined%
    \setlength{\unitlength}{252.28346457bp}%
    \ifx\svgscale\undefined%
      \relax%
    \else%
      \setlength{\unitlength}{\unitlength * \real{\svgscale}}%
    \fi%
  \else%
    \setlength{\unitlength}{\svgwidth}%
  \fi%
  \global\let\svgwidth\undefined%
  \global\let\svgscale\undefined%
  \makeatother%
  \begin{picture}(1,0.47191011)%
    \lineheight{1}%
    \setlength\tabcolsep{0pt}%
    \put(0,0){\includegraphics[width=\unitlength,page=1]{twtt_principle.pdf}}%
    \put(0.1794028,0.32518723){\color[rgb]{0,0,0}\makebox(0,0)[rt]{\lineheight{1.25}\smash{\begin{tabular}[t]{r}satellite B\end{tabular}}}}%
    \put(0.18073463,0.06676025){\color[rgb]{0,0,0}\makebox(0,0)[rt]{\lineheight{1.25}\smash{\begin{tabular}[t]{r}satellite A\end{tabular}}}}%
    \put(0.85196652,0.32754967){\color[rgb]{0,0,0}\makebox(0,0)[lt]{\lineheight{1.25}\smash{\begin{tabular}[t]{l}$\tau_{\text{B}}$\end{tabular}}}}%
    \put(0.85196652,0.0691227){\color[rgb]{0,0,0}\makebox(0,0)[lt]{\lineheight{1.25}\smash{\begin{tabular}[t]{l}$\tau_{\text{A}}$\end{tabular}}}}%
    \put(0.78642496,0.01123596){\color[rgb]{0,0,0}\makebox(0,0)[t]{\lineheight{1.25}\smash{\begin{tabular}[t]{c}$\tau_{\text{A},\text{RX},N}$\end{tabular}}}}%
    \put(0.24146869,0.28089888){\color[rgb]{0,0,0}\makebox(0,0)[t]{\lineheight{1.25}\smash{\begin{tabular}[t]{c}$\phi_{\text{B}}$\end{tabular}}}}%
    \put(0.28625085,0.1295412){\color[rgb]{0,0,0}\makebox(0,0)[t]{\lineheight{1.25}\smash{\begin{tabular}[t]{c}$\phi_{\text{A}}$\end{tabular}}}}%
    \put(0.46047789,0.43820225){\color[rgb]{0,0,0}\makebox(0,0)[t]{\lineheight{1.25}\smash{\begin{tabular}[t]{c}$T_{\text{ToF}}$\end{tabular}}}}%
    \put(0.68464592,0.36269663){\color[rgb]{0,0,0}\makebox(0,0)[t]{\lineheight{1.25}\smash{\begin{tabular}[t]{c}$\tau_{\text{B},\text{TX},N}$\end{tabular}}}}%
    \put(0.72948542,0.43820225){\color[rgb]{0,0,0}\makebox(0,0)[t]{\lineheight{1.25}\smash{\begin{tabular}[t]{c}$T_{\rm{ToF}}$\end{tabular}}}}%
    \put(0.59498168,0.43820225){\color[rgb]{0,0,0}\makebox(0,0)[t]{\lineheight{1.25}\smash{\begin{tabular}[t]{c}$T_{\rm{RESP}}$\end{tabular}}}}%
    \put(0.21348315,0.35955056){\color[rgb]{0,0,0}\makebox(0,0)[t]{\lineheight{1.25}\smash{\begin{tabular}[t]{c}$0$\end{tabular}}}}%
    \put(0.30337079,0.03402496){\color[rgb]{0,0,0}\makebox(0,0)[t]{\lineheight{1.25}\smash{\begin{tabular}[t]{c}$0$\end{tabular}}}}%
    \put(0.85196652,0.19113268){\color[rgb]{0,0,0}\makebox(0,0)[lt]{\lineheight{1.25}\smash{\begin{tabular}[t]{l}$t$\end{tabular}}}}%
    \put(0.26956779,0.2247191){\color[rgb]{0,0,0}\makebox(0,0)[t]{\lineheight{1.25}\smash{\begin{tabular}[t]{c}$0$\end{tabular}}}}%
    \put(0.50487064,0.36269663){\color[rgb]{0,0,0}\makebox(0,0)[t]{\lineheight{1.25}\smash{\begin{tabular}[t]{c}$\tau_{\text{B},\text{RX},N}$\end{tabular}}}}%
    \put(0.41507665,0.01123596){\color[rgb]{0,0,0}\makebox(0,0)[t]{\lineheight{1.25}\smash{\begin{tabular}[t]{c}$\tau_{\text{A},\text{TX},N}$\end{tabular}}}}%
  \end{picture}%
\endgroup%

%% file: fig/hdl_designs.pdf_tex
\begingroup%
  \makeatletter%
  \providecommand\color[2][]{%
    \errmessage{(Inkscape) Color is used for the text in Inkscape, but the package 'color.sty' is not loaded}%
    \renewcommand\color[2][]{}%
  }%
  \providecommand\transparent[1]{%
    \errmessage{(Inkscape) Transparency is used (non-zero) for the text in Inkscape, but the package 'transparent.sty' is not loaded}%
    \renewcommand\transparent[1]{}%
  }%
  \providecommand\rotatebox[2]{#2}%
  \newcommand*\fsize{\dimexpr\f@size pt\relax}%
  \newcommand*\lineheight[1]{\fontsize{\fsize}{#1\fsize}\selectfont}%
  \ifx\svgwidth\undefined%
    \setlength{\unitlength}{252.28346457bp}%
    \ifx\svgscale\undefined%
      \relax%
    \else%
      \setlength{\unitlength}{\unitlength * \real{\svgscale}}%
    \fi%
  \else%
    \setlength{\unitlength}{\svgwidth}%
  \fi%
  \global\let\svgwidth\undefined%
  \global\let\svgscale\undefined%
  \makeatother%
  \begin{picture}(1,0.4494382)%
    \lineheight{1}%
    \setlength\tabcolsep{0pt}%
    \put(0.10137815,0.1978207){\color[rgb]{0,0,0}\rotatebox{90}{\makebox(0,0)[t]{\lineheight{1.25}\smash{\begin{tabular}[t]{c}AXI Memory Interconnect\end{tabular}}}}}%
    \put(0,0){\includegraphics[width=\unitlength,page=1]{hdl_designs.pdf}}%
    \put(0.18018838,0.29817189){\color[rgb]{0,0,0}\rotatebox{90}{\makebox(0,0)[t]{\lineheight{1.25}\smash{\begin{tabular}[t]{c}TX DMA\end{tabular}}}}}%
    \put(0,0){\includegraphics[width=\unitlength,page=2]{hdl_designs.pdf}}%
    \put(0.18018838,0.09398349){\color[rgb]{0,0,0}\rotatebox{90}{\makebox(0,0)[t]{\lineheight{1.25}\smash{\begin{tabular}[t]{c}RX DMA\end{tabular}}}}}%
    \put(0.25571658,0.29829873){\color[rgb]{0,0,0}\rotatebox{90}{\makebox(0,0)[t]{\lineheight{1.25}\smash{\begin{tabular}[t]{c}TX Unpack\end{tabular}}}}}%
    \put(0,0){\includegraphics[width=\unitlength,page=3]{hdl_designs.pdf}}%
    \put(0.25571658,0.09483796){\color[rgb]{0,0,0}\rotatebox{90}{\makebox(0,0)[t]{\lineheight{1.25}\smash{\begin{tabular}[t]{c}RX Pack\end{tabular}}}}}%
    \put(0,0){\includegraphics[width=\unitlength,page=4]{hdl_designs.pdf}}%
    \put(0.33436827,0.29829873){\color[rgb]{0,0,0}\rotatebox{90}{\makebox(0,0)[t]{\lineheight{1.25}\smash{\begin{tabular}[t]{c}TX FIFO\end{tabular}}}}}%
    \put(0,0){\includegraphics[width=\unitlength,page=5]{hdl_designs.pdf}}%
    \put(0.33436827,0.09483796){\color[rgb]{0,0,0}\rotatebox{90}{\makebox(0,0)[t]{\lineheight{1.25}\smash{\begin{tabular}[t]{c}RX FIFO\end{tabular}}}}}%
    \put(0,0){\includegraphics[width=\unitlength,page=6]{hdl_designs.pdf}}%
    \put(0.41598489,0.1978207){\color[rgb]{0,0,0}\rotatebox{90}{\makebox(0,0)[t]{\lineheight{1.25}\smash{\begin{tabular}[t]{c}Timing Controller\end{tabular}}}}}%
    \put(0,0){\includegraphics[width=\unitlength,page=7]{hdl_designs.pdf}}%
    \put(0.64689096,0.18893495){\color[rgb]{0,0,0}\makebox(0,0)[t]{\lineheight{1.25}\smash{\begin{tabular}[t]{c}AD9361 Core\end{tabular}}}}%
    \put(0.98250125,0.23410538){\color[rgb]{0,0,0}\rotatebox{90}{\makebox(0,0)[lt]{\lineheight{1.25}\smash{\begin{tabular}[t]{l}AD9364\end{tabular}}}}}%
    \put(0,0){\includegraphics[width=\unitlength,page=8]{hdl_designs.pdf}}%
    \put(0.93269377,0.19761222){\color[rgb]{0,0,0}\rotatebox{90}{\makebox(0,0)[t]{\lineheight{1.25}\smash{\begin{tabular}[t]{c}LVDS\end{tabular}}}}}%
    \put(0,0){\includegraphics[width=\unitlength,page=9]{hdl_designs.pdf}}%
    \put(0.85376114,0.28721893){\color[rgb]{0,0,0}\rotatebox{90}{\makebox(0,0)[t]{\lineheight{1.25}\smash{\begin{tabular}[t]{c}TX Interf.\end{tabular}}}}}%
    \put(0,0){\includegraphics[width=\unitlength,page=10]{hdl_designs.pdf}}%
    \put(0.85376114,0.10800546){\color[rgb]{0,0,0}\rotatebox{90}{\makebox(0,0)[t]{\lineheight{1.25}\smash{\begin{tabular}[t]{c}RX Interf.\end{tabular}}}}}%
    \put(0,0){\includegraphics[width=\unitlength,page=11]{hdl_designs.pdf}}%
    \put(0.06007609,0.42696629){\color[rgb]{0.50196078,0.50196078,0.50196078}\makebox(0,0)[t]{\lineheight{1.25}\smash{\begin{tabular}[t]{c}$100 \text{MHz}$\end{tabular}}}}%
    \put(0,0){\includegraphics[width=\unitlength,page=12]{hdl_designs.pdf}}%
    \put(0.25842697,0.42696629){\color[rgb]{0.50196078,0.50196078,0.50196078}\makebox(0,0)[t]{\lineheight{1.25}\smash{\begin{tabular}[t]{c}$61.44 \text{MHz}$\end{tabular}}}}%
    \put(0.64044944,0.42696629){\color[rgb]{0.50196078,0.50196078,0.50196078}\makebox(0,0)[t]{\lineheight{1.25}\smash{\begin{tabular}[t]{c}$122.88 \text{MHz}$\end{tabular}}}}%
    \put(0,0){\includegraphics[width=\unitlength,page=13]{hdl_designs.pdf}}%
    \put(0.0439713,0.23410538){\color[rgb]{0,0,0}\rotatebox{90}{\makebox(0,0)[lt]{\lineheight{1.25}\smash{\begin{tabular}[t]{l}PS\end{tabular}}}}}%
    \put(0,0){\includegraphics[width=\unitlength,page=14]{hdl_designs.pdf}}%
    \put(0.62368376,0.30328612){\color[rgb]{0,0,0}\makebox(0,0)[t]{\lineheight{1.25}\smash{\begin{tabular}[t]{c}TX DSP\end{tabular}}}}%
    \put(0,0){\includegraphics[width=\unitlength,page=15]{hdl_designs.pdf}}%
    \put(0.62368376,0.07856703){\color[rgb]{0,0,0}\makebox(0,0)[t]{\lineheight{1.25}\smash{\begin{tabular}[t]{c}RX DSP\end{tabular}}}}%
    \put(0,0){\includegraphics[width=\unitlength,page=16]{hdl_designs.pdf}}%
  \end{picture}%
\endgroup%

%% file: fig/hdl_timing_controller.pdf_tex
\begingroup%
  \makeatletter%
  \providecommand\color[2][]{%
    \errmessage{(Inkscape) Color is used for the text in Inkscape, but the package 'color.sty' is not loaded}%
    \renewcommand\color[2][]{}%
  }%
  \providecommand\transparent[1]{%
    \errmessage{(Inkscape) Transparency is used (non-zero) for the text in Inkscape, but the package 'transparent.sty' is not loaded}%
    \renewcommand\transparent[1]{}%
  }%
  \providecommand\rotatebox[2]{#2}%
  \newcommand*\fsize{\dimexpr\f@size pt\relax}%
  \newcommand*\lineheight[1]{\fontsize{\fsize}{#1\fsize}\selectfont}%
  \ifx\svgwidth\undefined%
    \setlength{\unitlength}{246.61417323bp}%
    \ifx\svgscale\undefined%
      \relax%
    \else%
      \setlength{\unitlength}{\unitlength * \real{\svgscale}}%
    \fi%
  \else%
    \setlength{\unitlength}{\svgwidth}%
  \fi%
  \global\let\svgwidth\undefined%
  \global\let\svgscale\undefined%
  \makeatother%
  \begin{picture}(1,0.50574713)%
    \lineheight{1}%
    \setlength\tabcolsep{0pt}%
    \put(0,0){\includegraphics[width=\unitlength,page=1]{hdl_timing_controller.pdf}}%
    \put(0.05992465,0.42316576){\color[rgb]{0,0,0}\makebox(0,0)[lt]{\lineheight{1.25}\smash{\begin{tabular}[t]{l}\texttt{ts\_counter} (rf\_clk)\end{tabular}}}}%
    \put(0,0){\includegraphics[width=\unitlength,page=2]{hdl_timing_controller.pdf}}%
    \put(0.05992465,0.21078542){\color[rgb]{0,0,0}\makebox(0,0)[lt]{\lineheight{1.25}\smash{\begin{tabular}[t]{l}\texttt{ps\_interface} (\SI{100}{\mega\hertz})\end{tabular}}}}%
    \put(0,0){\includegraphics[width=\unitlength,page=3]{hdl_timing_controller.pdf}}%
    \put(0.51610285,0.42241965){\color[rgb]{0,0,0}\makebox(0,0)[lt]{\lineheight{1.25}\smash{\begin{tabular}[t]{l}\texttt{tx\_control}  (rf\_clk)\end{tabular}}}}%
    \put(0,0){\includegraphics[width=\unitlength,page=4]{hdl_timing_controller.pdf}}%
    \put(0.51610285,0.20953652){\color[rgb]{0,0,0}\makebox(0,0)[lt]{\lineheight{1.25}\smash{\begin{tabular}[t]{l}\texttt{rx\_control}  (rf\_clk)\end{tabular}}}}%
    \put(0.53130876,0.16162361){\color[rgb]{0,0,0}\makebox(0,0)[lt]{\lineheight{1.25}\smash{\begin{tabular}[t]{l}\texttt{ts\_curr}[63:0]\end{tabular}}}}%
    \put(0.53130876,0.11600578){\color[rgb]{0,0,0}\makebox(0,0)[lt]{\lineheight{1.25}\smash{\begin{tabular}[t]{l}\texttt{ts\_start}[63:0]\end{tabular}}}}%
    \put(0.53130876,0.07038796){\color[rgb]{0,0,0}\makebox(0,0)[lt]{\lineheight{1.25}\smash{\begin{tabular}[t]{l}\texttt{samples\_out}\end{tabular}}}}%
    \put(0.93158353,0.07038796){\color[rgb]{0,0,0}\makebox(0,0)[rt]{\lineheight{1.25}\smash{\begin{tabular}[t]{r}\texttt{samples\_in}\end{tabular}}}}%
    \put(0,0){\includegraphics[width=\unitlength,page=5]{hdl_timing_controller.pdf}}%
    \put(0.00918952,0.03346388){\color[rgb]{0,0,0}\rotatebox{90}{\makebox(0,0)[lt]{\lineheight{1.25}\smash{\begin{tabular}[t]{l}\texttt{adc\_in}\end{tabular}}}}}%
    \put(0.44299269,0.37376065){\color[rgb]{0,0,0}\makebox(0,0)[rt]{\lineheight{1.25}\smash{\begin{tabular}[t]{r}\texttt{c\_value}[63:0]\end{tabular}}}}%
    \put(0.07513058,0.37376066){\color[rgb]{0,0,0}\makebox(0,0)[lt]{\lineheight{1.25}\smash{\begin{tabular}[t]{l}\texttt{ctrl}\end{tabular}}}}%
    \put(0.44299273,0.1614452){\color[rgb]{0,0,0}\makebox(0,0)[rt]{\lineheight{1.25}\smash{\begin{tabular}[t]{r}\texttt{tx\_start}[63:0]\end{tabular}}}}%
    \put(0.44299273,0.11582736){\color[rgb]{0,0,0}\makebox(0,0)[rt]{\lineheight{1.25}\smash{\begin{tabular}[t]{r}\texttt{rx\_start}[63:0]\end{tabular}}}}%
    \put(0.44299272,0.06914718){\color[rgb]{0,0,0}\makebox(0,0)[rt]{\lineheight{1.25}\smash{\begin{tabular}[t]{r}\texttt{ctrl}[31:0]\end{tabular}}}}%
    \put(0,0){\includegraphics[width=\unitlength,page=6]{hdl_timing_controller.pdf}}%
    \put(0.5313088,0.37450676){\color[rgb]{0,0,0}\makebox(0,0)[lt]{\lineheight{1.25}\smash{\begin{tabular}[t]{l}\texttt{ts\_curr}[63:0]\end{tabular}}}}%
    \put(0.5313088,0.32888895){\color[rgb]{0,0,0}\makebox(0,0)[lt]{\lineheight{1.25}\smash{\begin{tabular}[t]{l}\texttt{ts\_start}[63:0]\end{tabular}}}}%
    \put(0.5313088,0.28327111){\color[rgb]{0,0,0}\makebox(0,0)[lt]{\lineheight{1.25}\smash{\begin{tabular}[t]{l}\texttt{samples\_in}\end{tabular}}}}%
    \put(0,0){\includegraphics[width=\unitlength,page=7]{hdl_timing_controller.pdf}}%
    \put(0.93158353,0.28385502){\color[rgb]{0,0,0}\makebox(0,0)[rt]{\lineheight{1.25}\smash{\begin{tabular}[t]{r}\texttt{samples\_out}\end{tabular}}}}%
    \put(0,0){\includegraphics[width=\unitlength,page=8]{hdl_timing_controller.pdf}}%
    \put(1.01278143,0.09428765){\color[rgb]{0,0,0}\rotatebox{90}{\makebox(0,0)[lt]{\lineheight{1.25}\smash{\begin{tabular}[t]{l}\texttt{d\_out}\end{tabular}}}}}%
    \put(1.01278143,0.32340655){\color[rgb]{0,0,0}\rotatebox{90}{\makebox(0,0)[lt]{\lineheight{1.25}\smash{\begin{tabular}[t]{l}\texttt{dac\_out}\end{tabular}}}}}%
    \put(0.07513058,0.16125056){\color[rgb]{0,0,0}\makebox(0,0)[lt]{\lineheight{1.25}\smash{\begin{tabular}[t]{l}\textit{AXI4\_LITE}\end{tabular}}}}%
    \put(0.00976423,0.30774553){\color[rgb]{0,0,0}\rotatebox{90}{\makebox(0,0)[lt]{\lineheight{1.25}\smash{\begin{tabular}[t]{l}\texttt{d\_in}\end{tabular}}}}}%
    \put(0,0){\includegraphics[width=\unitlength,page=9]{hdl_timing_controller.pdf}}%
    \put(0.03041188,0.4875){\color[rgb]{0,0,0}\makebox(0,0)[lt]{\lineheight{1.25}\smash{\begin{tabular}[t]{l}Timing Controller\end{tabular}}}}%
    \put(0,0){\includegraphics[width=\unitlength,page=10]{hdl_timing_controller.pdf}}%
    \put(0.00939118,0.1861387){\color[rgb]{0,0,0}\rotatebox{90}{\makebox(0,0)[lt]{\lineheight{1.25}\smash{\begin{tabular}[t]{l}\textit{AXI}\end{tabular}}}}}%
  \end{picture}%
\endgroup%